# CMOS-Memristor Dendrite Threshold Circuits

Askhat Zhanbossinov, Kamilya Smagulova, and Alex Pappachen James
Bioinspired Microelectronics Systems Group,
School of Engineering, Nazarbayev University, Astana
apj@ieee.org; www.biomicrosystems.info

*Abstract*— Non-linear neuron models overcomes the limitations of linear binary models of neurons that have the inability to compute linearly non-separable functions such as XOR. While several biologically plausible models based on dendrite thresholds are reported in the previous studies, the hardware implementation of such non-linear neuron models remain as an open problem. In this paper, we propose a circuit design for implementing logical dendrite non-linearity response of dendrite spike and saturation types. The proposed dendrite cells are used to build XOR circuit and intensity detection circuit that consists of different combinations of dendrite cells with saturating and spiking responses. The dendrite cells are designed using a set of memristors, Zener diodes, and CMOS NOT gates. The circuits are designed, analyzed and verified on circuit boards.

*Index Terms*— Memristors, Dendrite models, Neural Circuits, Analog Circuits, Gates, Image processing

## I. INTRODUCTION

Biological neural networks can perform intelligent computing task following a hierarchical, modular and sparse processing of signals [1], [2]. As opposed to the non-linear neuron models that can be used in a wide range of applications [3]–[5], the earlier neuron models were considered to have linear characteristics that allows to perform only several basic functions [6], [7]. There are several theoretical linear and non-linear neuron models proposed in recent years [2], [8]. However, the hardware implementations of non-linear dendrite neuron model remains as an open challenge. Expanding the computational capacity of a single neuron, [9] proposes a biologically relevant non-linear model of neuron that allows to implement linearly non-separable Boolean functions. In this paper, we propose a hardware implementation of XOR function based on non-linear neuron model illustrated in [9] as well as the application of non-linear neuron model for pixel intensity detection for possible use in color (or intensity) segmentation and edge detection in images. The proposed XOR function and pixel intensity detector circuit implementation based on non-linear neuron model consists of a combination of saturating and spiking dendrites response circuits.

## II. BACKGROUND

Previously, it is believed that dendrites in neuron only function as a provider of signals where action potential doesn't take place [10]. Discovery of various voltage-gated ion channels: sodium, potassium and calcium channels in dendrites; revealed more complex structure and functionality of the dendrites than it is previously thought [11]. Neuron has several mechanisms to intensify weak input signals in dendrites. One through spatial summation of synaptic inputs, second through placement of high density of voltage-gated sodium and calcium ion channels [12].

There are two distinctive regions in dendrite that receive synaptic inputs performant path (PP) placed 500-750$\mu$m from soma and Schaffer-collateral (SC) path 250-500$\mu$m from soma, PP stimulates spike formation and SC decides whether signal will be passed to the soma region [13]. Here SC region acts as threshold unit, although it can be with or without potassium channels, which in its turn also acts as threshold unit which regulates over excitement of the synaptic events in synaptic regions and inhibits action potential in dendrites by moving cations outwards [14]. Next aspect of neuron other than signal stimulation, propagation control is ability to learn and memorize. One of the suggested mechanisms of memorization and learning is though actin based plasticity of dendrites [15]. In other words, memorization and learning processes maintained through grow of actins in apical region of dendrites, forcing dendrites to hold certain position [10].

Those advances in understanding of neuron physiology gives some suggestions about what architecture might be viable for artificial neuron design [1]. There are several points that we need to consider to be able to create artificial neuron. Point one, we need to maintain multiple inputs as there are multiple dendrites in neuron, second, we need to have threshold units in dendrite regions as well as in soma region, third, algorithm functioning in multiple input and single output environment has to be established.

Recently, it was found that linearly non-separable functions, as XOR, can be constructed by a binary neuron model with a single non-linear dendrite [9]. In [9], it is suggested that dendrites are capable of computing linearly non-separable functions as Boolean mathematics, and the best way to maintain functionality as in biological counterpart is using Exclusive OR function. Here, we present functional design of dendritic threshold logic neuron model closely resembling biological counterpart [1].

## III. PROPOSED CIRCUITS

In comparison to linear binary models of neurons that cannot compute linearly non-separable functions [6], [7], non-linear neuron models allow to implement Boolean functions due to the ability of non-linearly changing the stage from active to inactive and vice versa [3]. Fig. 1 illustrates the dendrite spike and dendrite saturation functions. The dendrite saturation sub-units are strictly sub-linear, while the dendrite spiking sub-units are both supra linear and sub-linear. Dendrite spiking and saturation functions are characterized by the threshold value $\theta$ and the maximum output height $h$ [9], [1].

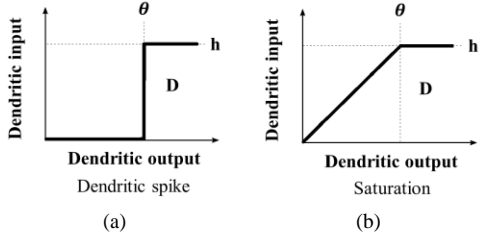

Fig. 1. Logical dendritic non-linearities: (a) dendritic spike type and (b) saturation type [1]

Fig. 2 represents the proposed CMOS-memristive dendrite cells. It consists of a Zener diode, memristor and CMOS inverter circuits. The Zener diode reverse breakdown voltage defines the maximum output height $h$, and threshold value $\theta$ is defined by the memristor-inverter combination for circuit Fig. 2(a), while $\theta$ is defined by the breakdown voltage of Zener for circuit Fig. 2(b). The example of a measured output characteristic of the proposed cells is shown in Fig. 3, Zener breakdown at 4.2V, limiting the height $h$.

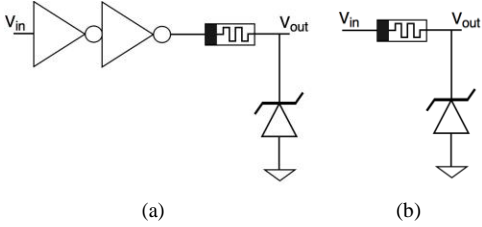

Fig. 2. CMOS-memristor dendrite threshold logic cell. Equivalent circuit of (a) dendrite spike type and (b) saturation type.

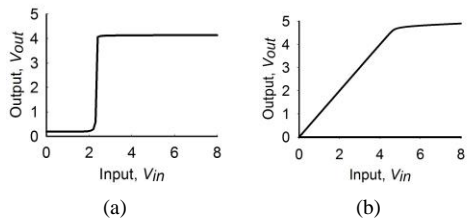

Fig. 3. The response of circuits in Fig. 2, (a) dendritic spike type and (b) saturation type.

### A. XOR Circuit

In this section dendrite spike type and saturation type neuron models are used in combination to build XOR gate. Apart from being a logical gate, XOR as a function is a linearly non-separable function, that can be a basis of simple classification problems. The proposed two staged neuron model based circuit implementation of XOR Boolean function is shown on Fig. 4. Fig. 4 (a) is the neuron model for the XOR functionality which has been proposed in the work [9] and (b) is its circuit implementation.

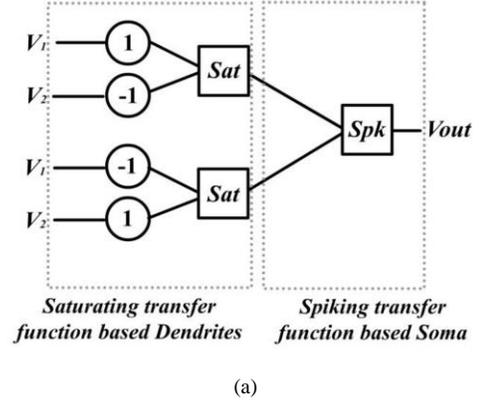

(a)

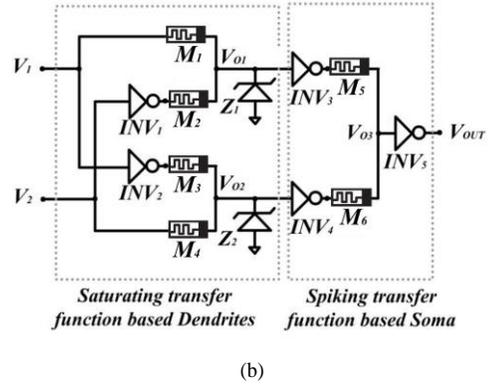

(b)

Fig. 4. Two staged neuron model of XOR functionality, where (a) is the neuron model and (b) its circuit implementation. In (a) the values given in circle are the synaptic weights, Sat is the saturating transfer function based threshold unit and Spk is the spiking transfer function based threshold unit.

The inverters $INV_1$ and $INV_2$ are used to implement the -1 synaptic weight as mentioned in neuron model. It is working based on the function $F_1$ given in the Eq. 1, where $\theta_1$ is the inverter threshold and $x$ is the input.

$$F_1(x) = \begin{cases} 1 & \text{if } x < \theta_1 \\ -1 & \text{if } x \leq \theta_1 \end{cases} \quad (1)$$

The saturating transfer function $F_{sat}$, as mentioned in [16], is given in Eq. 2, where $\theta_2$ is the dendrite threshold and $a$ is the input. The threshold unit based on this function is implemented using a Zener diode.

$$F_{sat}(a) = \begin{cases} 1 & \text{if } a \geq \theta_2 \\ \frac{a}{\theta_z} & \text{otherwise} \end{cases} \quad (2)$$

While using Zener diode, Eq. 2 will changes to Eq. 3, where $\theta_2$ is the peak inverse voltage of Zener diode and $a$ is the input.

$$F_{sat}(a) = \begin{cases} \theta_2 & \text{if } a \geq \theta_2 \\ a & \text{otherwise} \end{cases} \quad (3)$$

Inverters $INV_3$, $INV_4$ and $INV_5$ are the spiking transfer function [16] based threshold units. Here $INV_3$ and $INV_4$

works based on the Eq. 4 and $INV_5$ based on the Eq. 5. In Eq. 4, $\theta_2$ refers to the peak inverse voltage of Zener diode and $b$ is the input. In Eq. 5, $c$ is the input and $\theta_3$ is the threshold voltage of the inverter and it should satisfy the condition $max(F_{spk1}(b)) > \theta_3 > max(F_{spk1}(b))/2$.

$$F_{spk1}(b) = \begin{cases} 0 & \text{if } b = \theta_2 \\ 1 & \text{otherwise} \end{cases} \quad (4)$$

$$F_{spk2}(c) = \begin{cases} 0 & \text{if } c = \theta_3 \\ 1 & \text{otherwise} \end{cases} \quad (5)$$

The threshold of inverters $INV_3$ and $INV_4$ should be selected in accordance to the Zener break down voltage ($\theta_2$) in such a way that it should be less than $\theta_2$ but close to it. The circuit was implemented using chalcogenide based ion-conducting memristor, transistors (0.18$\mu$ MOS technology), and Zener diodes. The measured transients in Fig 5 validates the working of the implemented XOR circuit shown in Fig. 4.

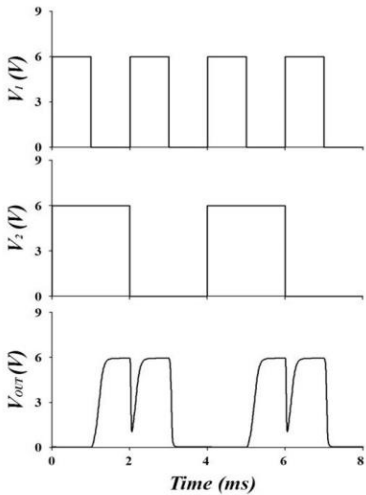

Fig. 5. The input and output signal wave-forms shows the operational response of the proposed XOR circuit in Fig. 4.

### B. Pixel Intensity Detection Circuit

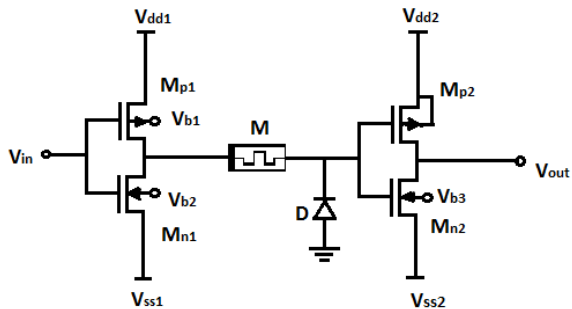

Fig. 6. Multiple threshold circuit for detecting image edges.

Detection of binary edges from an image requires atleast two intensity threshold operations around the spatially continuous edge regions in an object. The aim of this section to create a pixel intensity detection circuit that responds only to a range of intensity voltages read from a CMOS sensor. The circuit in this case would have a low threshold that rejects the low voltage values, and a higher threshold that rejects the higher voltage values. Applying the input pixel voltages to such a circuit, it could be observed that lower the difference between the low threshold and high threshold indicates a thinner would be the edge in a high contrast image. For low contrast images, the same threshold levels can be used to detect regions and hence will be used for image segmentation. The nature of the image, and the threshold levels would determine the application and use of this intensity detection circuit. Fig. 6 shows the pixel intensity detection circuit constructed with memristor (chalcogenide based ion-conducting), transistors (0.18$\mu$ MOS technology), and Zener diode. The NOT gate that acts like a variable threshold circuit ($M_{p1}$ and $M_{n1}$) followed by saturation threshold logic circuit (memristor and Zener) forms the inverted dendrite spike logic as the first threshold operator. The second threshold operation is realized by the NOT gate ($M_{p2}$ and $M_{n2}$). Fig. 7 shows the output response ($V_{out1}$ and $V_{out2}$) for different threshold configurations for the circuit Fig. 6. The response $V_{out1}$ was obtained when supply voltages $V_{dd1}$, $V_{dd2}$, $V_{ss1}$ are equal to 1.6V whereas $V_{ss2}$ is 0.6V. Bulk voltages of transistors $M_{p1}$, $M_{n1}$ and $M_{n2}$ are 1 V. The second response $V_{out2}$ was obtained by changing supply voltages $V_{dd1}$ and $V_{ss1}$ to 1.9V and $V_{dd2}$ to 2.6V. Figure 8, shows an illustrative example of applying the response functions in Fig. 7 to a Gaussian image. The increased width and height of the response function in Fig 7 is translated to thicker and brighter image circle in Fig 8.

The practical implementation of this circuit requires parallel application of the pixel outputs to be fed as input to the circuit in Fig. 6. For every range of threshold responses tuned, a different color (range of intensities) region in the image will be obtained. This is useful for detecting edges in the images, and color segmentation of regions. Some applications of color segmentation are in face detection, region detection in medical images, object tracking and industrial vision systems.

## IV. CONCLUSION

In this paper, we presented circuit designs for CMOS-memristive dendrite cells for mimicking dentrite spike and saturation type responses. These basic designs were further extended to develop XOR gate and a pixel intensity detection circuits. The XOR implementation signifies the neural circuit ability to be useful for a simple non-separable binary classification problem, to be useful as a logical gate, and also as a binary comparator circuit. The pixel detector circuit signify the application of the neural models to be integrated into pixel sensors to detect intensity levels (mapped as voltage levels) for real-time practical applications of color segmentation and edge detection. The proposed design can be extended to build image recognition systems for object detection and tracking, among other applications. This requires us to build array circuits for the neurons and relevant circuit analysis in benchmarking; which is left for future work.

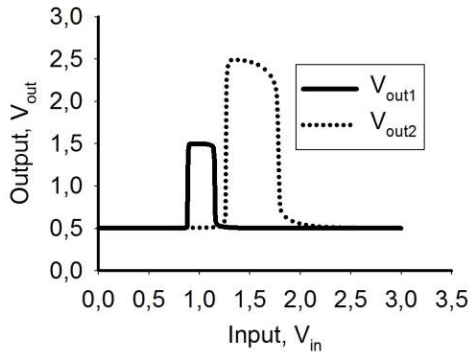

Fig. 7. Transfer characteristics to validate the working of the proposed circuit in Fig. 6 for generating multiple threshold.

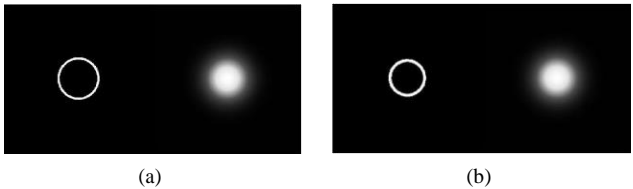

(a)    (b)

Fig. 8. The output images in the left side of the images (a) and (b) to that of the Gaussian image in the right side for the transfer characteristics in Fig 7, where the $V_{out1}$ and $V_{out2}$ responses corresponds to (a) and (b) respectively.

ACKNOWLEDGEMENTS

We would like to acknowledge the discussions with Dr Romain Cazé, Imperial College London on the biological aspects of the dendrite model; Ms Olga Krestinskaya for verifying the results. Also, we acknowledge the NU research funding provided for carrying out this work.